\documentclass[%
 reprint,
superscriptaddress,
 amsmath,amssymb,
 aps,
prb,
citeautoscript,
]{revtex4-2}

\usepackage{graphicx}
\usepackage{dcolumn}
\usepackage{bm}
\usepackage{hyperref}
\usepackage{siunitx}
\usepackage{braket}
\usepackage{tikz}
\usetikzlibrary{quantikz2}
\usepackage{floatrow}
\usepackage[caption=false]{subfig} 
\usepackage{multirow}

\usepackage{csquotes}
\usepackage{booktabs}
\usepackage{orcidlink}

\begin{document} 

\title{Multireference error mitigation for quantum computation of chemistry}

\author{Hang Zou}
	\affiliation{
	Department of Chemistry and Chemical Engineering, 
	Chalmers University of Technology, 41296 Gothenburg, Sweden
    }
    \affiliation{Department of Computer Science and Engineering, Chalmers University of Technology and University of Gothenburg, 41296 Gothenburg, Sweden 
    }

\author{Erika Magnusson \orcidlink{0009-0002-1948-487X}}
	\affiliation{
	Department of Chemistry and Chemical Engineering, 
	Chalmers University of Technology, 41296 Gothenburg, Sweden
    }
    
\author{Hampus Brunander}
	\affiliation{
	Department of Chemistry and Chemical Engineering, 
	Chalmers University of Technology, 41296 Gothenburg, Sweden
}

\author{Werner Dobrautz \orcidlink{0000-0001-6479-1874}}
 \email{werner.dobrautz@gmail.com}
         \affiliation{
        Center for Advanced Systems Understanding, 
        Helmholtz-Zentrum Dresden-Rossendorf, Germany
}
        \affiliation{
        Center for Scalable Data Analytics and Artificial Intelligence Dresden/Leipzig, 
        TU Dresden, Germany
}
 	\affiliation{
 	Department of Chemistry and Chemical Engineering, 
 	Chalmers University of Technology, 41296 Gothenburg, Sweden
}

\author{Martin Rahm \orcidlink{0000-0001-7645-5923}}
 \email{martin.rahm@chalmers.se}
	\affiliation{
	Department of Chemistry and Chemical Engineering, 
	Chalmers University of Technology, 41296 Gothenburg, Sweden
}

\begin{abstract}
Quantum error mitigation (QEM) strategies are essential for improving the precision and reliability of quantum chemistry algorithms on noisy intermediate-scale quantum devices.	
Reference-state error mitigation (REM) is a cost-effective chemistry-inspired QEM method that performs exceptionally well for weakly correlated problems. 
However, the effectiveness of REM is often limited when applied to strongly correlated systems. 
Here, we introduce multireference-state error mitigation (MREM), an extension of REM that systematically captures quantum hardware noise in strongly correlated ground states by utilizing multireference states.
A pivotal aspect of MREM is using Givens rotations to efficiently construct quantum circuits to generate multireference states. 
To strike a balance between circuit expressivity and noise sensitivity, we employ compact wavefunctions composed a few dominant Slater determinants. 
These truncated multireference states, engineered to exhibit substantial overlap with the target ground state, can effectively enhance error mitigation in variational quantum eigensolver experiments.	
We demonstrate the effectiveness of MREM through comprehensive simulations of molecular systems $\mathrm{H_2O, ~N_2, ~and ~F_2}$, underscoring its ability to realize significant improvements in computational accuracy compared to the original REM method. 
MREM broadens the scope of error mitigation to encompass a wider variety of molecular systems, including those exhibiting pronounced electron correlation.
\end{abstract}

\maketitle

\section{Introduction}
Quantum computers hold considerable promise for solving computationally infeasible problems for classical computers \cite{1,2,3}.  They have the potential to speed up the simulation of quantum systems and to offer exponential memory storage capabilities \cite{4,5,6}.
Quantum chemistry, in particular, is expected to gain potential long-term benefits from advances in quantum computing \cite{7,8,9,10,11}. However, current noisy intermediate-scale quantum (NISQ) devices \cite{13} are susceptible to noise, which can result in loss of coherence during computation, thus undermining potential quantum advantages \cite{12, yan2023limitationsnoisyquantumdevices}. Even for NISQ algorithms featuring shallow circuits, such as the variational quantum eigensolver (VQE) \cite{27,28} or variational quantum imaginary time evolution (VarQITE) \cite{McArdle2019, Yuan2019, Sokolov2023}, errors inevitably accumulate during computation, leading to unreliable results. The current number and fidelity of physical qubits do not meet the demands of fault-tolerant quantum computing utilizing quantum error-correcting codes \cite{15,14,16,17}. Therefore, pursuing alternative approaches to achieve meaningful results and accelerate the practical application of NISQ devices is crucial.  

Research in quantum error mitigation (QEM) shifts the focus from hardware resources to sophisticated information processing techniques \cite{18,19,20,21,22,23,24,25,29,30,31,32,33,34,35,36, Skogh2024}. QEM typically involves executing an ensemble of noisy circuits multiple times or making moderate circuit modifications, followed by post-processing the noisy data to infer ideal computational results. Numerous QEM methods have been proposed to improve the quality of results calculated with NISQ hardware, including error extrapolation \cite{20,32}, probabilistic error cancellation \cite{20,33}, virtual distillation \cite{24,34}, measurement error mitigation \cite{29,30}, symmetry constraints \cite{23, Skogh2024}, subspace expansions \cite{31,36}, learning-based methods \cite{25,35}, and reference-state error mitigation (REM) \cite{22}. The cost of using a QEM strategy is paid in additional sampling costs, which primarily determine the feasibility and scalability of a QEM protocol. Many QEM methods incur exponential sampling overhead as circuit depth and qubit count increase. 

Several universal frameworks have been proposed to evaluate the minimum sampling requirements of general QEM protocols, highlighting the inherent exponential challenges to QEM scalability \cite{37,38,39,86}. These frameworks provide task-agnostic guarantees, assuming no prior knowledge about the problem structure. However, in specific domains such as quantum chemistry, physically motivated assumptions -- for example, the availability of a good trial wavefunction or an approximate model of the target state -- can often be leveraged to design more efficient QEM strategies, significantly reducing the sampling cost in practice.

The REM method, described in detail in Sec.~\ref{section2b}, leverages chemical insight to provide a low-complexity error mitigation approach, requiring at most one additional algorithm, e.g. VQE/VarQITE, iteration \cite{22, 55}. The idea of REM is to mitigate the energy error of a noisy target state measured on a quantum device by first quantifying the effect of noise on a close-lying reference state. The reference state, often also set to be the initial state of the calculation, is chosen to be (i) exactly solvable on a classical computer and (ii) practical to prepare and measure on a quantum device. The cost of implementing REM is solely attributed to the preparation of the reference state on a quantum device and the determination of its exact/noiseless energy using a classical computer. Provided that the reference state is also the initial state, there is no need for additional measurements of the reference state's energy on a quantum device.

The first work on REM demonstrated the use of a single-reference Hartree-Fock (HF) state to achieve significant error mitigation gains \cite{22}. 
The HF state, described as an \enquote{uncorrelated} single determinant, can be easily prepared on a quantum computer using only Pauli-X gates. The circuits for HF state preparation maintain a constant complexity and are Clifford circuits, which can be efficiently simulated classically, as stated by the Gottesman-Knill theorem \cite{46}. The HF state serves as the starting point for many wavefunction theories and ensures sufficient overlap with the target ground state in most molecules \cite{27,28,54}. The effectiveness of using an HF reference for REM has subsequently been repeatedly demonstrated, \textit{e.g.}, in Refs.~\cite{22,55,56}. In contrast, random references generated from Clifford groups are almost guaranteed to be ineffective. 
The REM method combined with single-reference states such as HF nearly establishes a lower bound on QEM costs for quantum chemistry applications, as it incurs only the classical computational cost of a trivial state. 

Although REM has proven effective in weakly correlated systems, its utility is more limited in the presence of strong electron correlation, such as in bond-stretching regions \cite{22}. 
This limitation arises because REM assumes that the chosen reference state -- typically a single Slater determinant (\textit{e.g.}, Hartree-Fock) -- is a reasonable approximation of the target ground state.
However, in systems with strong correlation, the exact wavefunction often takes the form of a \textit{multireference (MR) state}, i.e., a linear combination of multiple Slater determinants (SDs) with similar weights. 
In such cases, a single determinant no longer provides sufficient overlap with the true ground state, and using it as a reference leads to inaccurate error mitigation.
Consequently, REM becomes unreliable for these problems, motivating the need for an extended framework that incorporates multiconfigurational states with better overlap to the correlated target wavefunction.

In this work, we address this limitation by introducing multireference-state error mitigation (MREM), an extension of REM that systematically incorporates MR states into the error mitigation protocol.
MREM uses approximate MR wavefunctions generated by inexpensive conventional methods and prepares them on quantum hardware using using physically motivated, symmetry-preserving quantum circuits. 
In particular, we employ Givens rotations to construct multireference states with controlled expressivity and efficient hardware implementation.

While Givens rotations are central to our implementation, they are not the only possible method for MR state preparation.
Alternative strategies include low-depth adaptive ansätze \cite{grimsley_adaptive_2019, 80}, adiabatic state preparation \cite{aspuru2005simulated, veis2014adiabatic}, or non-orthogonal subspace methods \cite{huggins2020non,baek2023say,85}.
However, these often lack symmetry guarantees, require extensive parameter tuning, or introduce circuit design and measurement complexity.
In contrast, Givens rotations offer a structured and physically interpretable approach to building linear combinations of SDs from a single reference configuration.
They preserve key symmetries such as particle number and spin projection, and are known to be universal for quantum chemistry state preparation tasks \cite{42}.
These features, along with widespread prior use in constructing symmetry-adapted ansätze \cite{40,41,42,70,74}, make Givens-based circuits a compelling and efficient choice for implementing MREM (see Sec.~\ref{section2d}).

To avoid confusion, we want to note that the \enquote{multireference}-states used in MREM are not necessarily obtained through conventional quantum chemistry \textit{multireference} methods. 
Rather, they refer to truncated multi-determinant wavefunctions,\textit{i.e.}, a linear combination of SDs, derived from various classical methods including both single- and multireference approaches, as detailed below.

The rest of this paper is organized as follows. In Sec. \ref{section2}, we provide the basic concepts of VQE, an overview of the original REM, the basics of Givens rotations, and the methodology for realizing MREM using Givens rotations. Sec. \ref{section3} outlines computational details. In Sec. \ref{section4}, we demonstrate the performance improvements of MREM compared to single-reference REM for the molecular systems H$_2$O, N$_2$, and F$_2$. Finally, Sec. \ref{section5} offers our conclusions and perspectives on the future directions of the MREM method.

\section{Theory \label{section2}}
\subsection{The variational quantum eigensolver}
While REM and our extension MREM do not rely on any specific variational algorithm, we have chosen to demonstrate the approach in the framework of VQE -- perhaps the most well-known variational quantum algorithm -- for familiarity.
The electronic Hamiltonian $\hat{H}$ of a molecular system can be expressed in second quantization as  
\begin{equation}\label{eq:ferm-ham}
	\hat{H}=\sum_{pq} h_p^q \hat a_p^\dagger \hat a_q+\frac12\sum_{pqrs} g^{rs}_{pq} \hat a_p^\dagger \hat a_q^\dagger \hat a_s \hat a_r,  
\end{equation}
where $h_p^q$ and $g_{pq}^{rs}$ represent the one- and two-electron integrals, and $\hat a^{(\dagger)}_p$ represent the fermionic annihilation (creation) operators in spin-orbital $p$.  
In quantum computing, a fermion-to-qubit mapping such as the Jordan-Wigner (JW) \cite{43} or Bravyi-Kitaev \cite{44} transformation is required to convert the fermionic Hamiltonian, Eq.~\eqref{eq:ferm-ham}, into a qubit Hamiltonian, expressed as a sum of $N$-qubit Pauli operators $\hat P_\alpha$:
\begin{equation}
	\hat{H}=\sum_\alpha h_\alpha \hat P_\alpha,
\end{equation}
with coefficients $h_\alpha$.
The VQE algorithm aims to find $E(\boldsymbol{\theta})$, an approximation to the ground state energy $E_0$ dependent on circuit parameters $\boldsymbol{\theta}$ such that
\begin{equation}
	E_{0}\leq  E (\boldsymbol{\theta}) = \min_{\boldsymbol{\theta}}\langle\psi(\boldsymbol{\theta})|\hat{H}|\psi(\boldsymbol{\theta})\rangle .
\end{equation}
We employ an ansatz, a parameterized quantum circuit $\hat U(\boldsymbol{\theta})$ to prepare a trial quantum state 
$\left|\psi(\boldsymbol{\theta})\right\rangle = \hat U(\boldsymbol{\theta})\left|\psi_0\right\rangle$, and calculate the energy expectation value from many measurements. 
The number of measurements (shots) for energy estimation using the molecular Hamiltonian scales as $\mathcal{O}\left({N^4}/{\epsilon^2}\right)$, where $\epsilon$ is the desired precision and $N$ is the number of qubits. 
The ansatz structure enables a quantum computer to explore a wide range of quantum states within a constrained expressible subspace, with optimization of its parameters guiding the search within this space. The state $|\psi_0\rangle=\hat U_{\mathrm{init}}\ket{0}$ is an initial state that, optimally, has a large overlap with the ground state. 

\subsection{Reference-state error mitigation \label{section2b}}
Noise in the quantum system disrupts state preparation and measurements in quantum algorithms, like the VQE, limiting the ansatz's accessible space. 
Consequently, the energy estimate from the noisy VQE will be significantly higher than the true ground state energy. 
REM can effectively mitigate VQE energy errors by capturing the energy error caused by noise in a well-chosen reference state.  

The procedure of REM, here exemplified for VQE, is as follows:
\begin{enumerate}
	\item Select a reference state $\ket{\psi_{\mathrm{ref}}}$ and determine its exact/noiseless energy $E_\text{ex}(\boldsymbol{\theta}_{\mathrm{ref}})$ using a classical computer.
	\item  Prepare the reference state on a quantum computer and measure the noisy energy $E_{\mathrm{VQE}}(\boldsymbol{\theta}_{\mathrm{ref}})$.   The effect of noise on the reference state is quantified as $\Delta E_{\mathrm{REM}}=E_{\mathrm{VQE}}(\boldsymbol{\theta}_{\mathrm{ref}})-E_{\mathrm{ex}}(\boldsymbol{\theta}_{\mathrm{ref}})$.
	\item Perform the VQE algorithm on the same circuit as Step 2 to obtain the noisy energy expectation value $E_{\mathrm{VQE}}(\boldsymbol{\theta}_{\mathrm{min}, \mathrm{VQE}})$.
	\item Apply the REM correction to obtain an error-mitigated result: \\$
	E_{\mathrm{REM}}(\boldsymbol{\theta}_{\min , \mathrm{VQE}})=E_{\mathrm{VQE}}(\boldsymbol{\theta}_{\min , \mathrm{VQE}})-\Delta E_{\mathrm{REM}}
	.$ 
\end{enumerate}

A good first choice for a single-reference state is the HF state: 
\begin{equation} 
	|\psi_{\mathrm{HF}}\rangle= \hat a_{p_1}^\dagger \hat a_{p_2}^\dagger \cdots \hat a_{p_{n_e}}^\dagger|0\rangle^{\otimes 2m}=\ket{0\cdots 01\cdots 11},
\end{equation} 
where $n_e$ denotes the number of electrons, and $2m$ is the total number of spin-orbitals. In this work, we chose an interleaved spin ordering of electrons for a Slater determinant. 
For instance, a generic singlet ($S_z=0$) HF state is represented as $\ket{0\cdots 01\cdots 11}$, \textit{i.e.}, $\ket{\cdots n^\alpha_{n_\alpha+1} n^\beta_{n_\beta}\cdots n^\beta_1n^\alpha_1}$, where the rightmost qubit refers to the first qubit, and $n_i^{\alpha, \beta}\in \{0, 1\}$ represent the occupation number of spin-orbitals with electron spins $\alpha$ or $\beta$. 
Here, $n_\alpha=n_\beta=n_e/2$ indicate the number of each spin for the $S_z=0$ state. 
In the JW mapping, each qubit corresponds to a spin-orbital of the molecular system, where the $\ket{0}$ and $\ket{1}$ states locally encode the occupation of each spin-orbital.  

\subsection{Multireference-state error mitigation}  
As outlined previously, the REM framework is general and can accommodate any reference state that is both practical to prepare on a quantum device and exactly solvable on a classical computer.
The multireference-state error mitigation (MREM) strategy presented here constitutes a specific instantiation of REM, in which the reference state is a multiconfigurational wavefunction.
The primary distinction between MREM and earlier single-reference REM implementations lies in Steps 1 and 2 of the general procedure (see Sec.~\ref{section2b}).
These steps involve the selection and preparation of a multireference (MR) state -- a superposition of multiple Slater determinants -- to better approximate the correlated electronic structure of the system:
\begin{equation}
	|\psi_{\mathrm{MR}}\rangle=\sum_{j}c_{j}\left|n_N\cdots n_2 n_1\right\rangle.
\end{equation}

\begin{figure*}
	\centering
	\includegraphics[width=\linewidth]{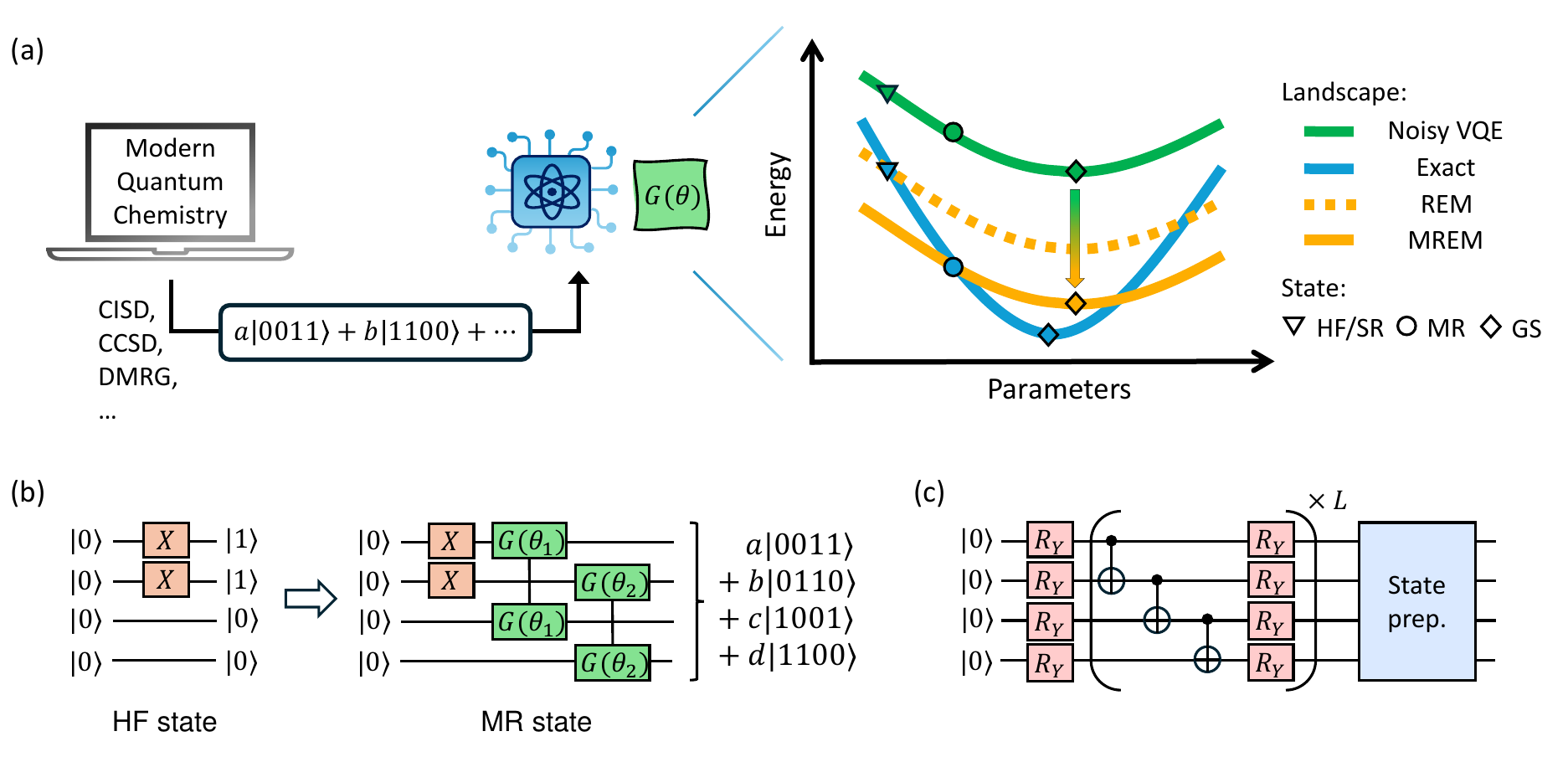}
	\caption{Givens-based MREM method description. (a) The workflow begins with conventional quantum chemistry methods (\textit{e.g.}, CISD, CCSD, DMRG) evaluating MR states, which are subsequently prepared on quantum hardware through Givens rotations. This MREM approach provides enhanced error mitigation capabilities compared to the single-reference REM methods, as illustrated in the energy landscape diagram where the MREM solutions (orange solid line) better approximate the exact solution (blue line) compared to the single-reference approaches (orange dashed line).	
		(b) A MR state preparation circuit: parameterized Givens rotations generate excited configurations from an HF state. 
		(c) A hardware-efficient ansatz with $R_Y$ single-qubit rotations and linear connectivity of CNOT gates. The ansatz is placed before the state preparation circuit, with all parameters initialized to 0.}
	\label{fig1}
\end{figure*}

As illustrated in Fig.~\ref{fig1}~(a), the steps of MREM are:
\begin{enumerate}
	\item Select a suitable (possibly truncated) MR state guided by conventional quantum chemistry theories and determine its exact/noiseless energy with affordable cost on a classical computer.
	\item Prepare the MR state on quantum hardware (here using Givens rotations, see Sec.~\ref{section2d}) and measure its noisy energy, obtaining the MREM error estimation, $\Delta E_{\rm MREM}$.
	\item Apply Steps 3 and 4 from the general REM procedure. The noisy ground state energy measured on the quantum computer, combined with the energy correction provided by the MR state, yields the error-mitigated ground state energy.
\end{enumerate}

Our Givens-based MREM implementation is designed to flexibly incorporate MR states derived from a range of conventional quantum chemistry methods that capture electron correlation.
In this work, we demonstrate this flexibility using MR states obtained from post-Hartree–Fock single-reference methods such as configuration interaction with singles and doubles (CISD) \cite{53} and coupled cluster with singles and doubles (CCSD) \cite{51}, as well as multireference methods like the density matrix renormalization group (DMRG) \cite{50}.

While this versatility is a strength of the approach, it also introduces practical challenges in implementation.
A key concern is the efficient preparation of the chosen MR states on quantum hardware without incurring excessive circuit complexity.
If state preparation becomes too costly in terms of gate depth or non-Clifford resources, the additional noise can negate the benefits of improved expressivity.
To address this trade-off, we employ truncated wavefunctions that retain only a small number of significant configurations -- 2–3 SDs in this work -- aligning closely with the principles of selected configuration interaction \cite{Huron1973, Garniron2018} or full configuration interaction quantum Monte Carlo \cite{Booth2009, Guther2020, Dobrautz2021, Dobrautz2019}.
This simplification strikes a practical balance between error mitigation performance and the cost of additional circuit depth and noise. 

\subsection{Givens rotations  \label{section2d}}
Givens rotations provide a computationally efficient framework for implementing the transformations required in the preparation of MR states within quantum circuits. 
In our approach, MR states are constructed by applying a sequence of Givens rotations to a single-reference determinant, systematically introducing electronic excitations to generate coherent superpositions of multiple Slater determinants. 

Mathematically, a Givens rotation operates through a unitary transformation matrix that acts on two selected elements while leaving others unchanged. The fundamental block matrix takes the form 
\begin{equation}
	U(\theta)= \left(\begin{array}{cc} \cos(\theta/2) & -\sin(\theta/2) \\ \sin(\theta/2) &\cos(\theta/2) \end{array}\right).
	\label{eq:rotmatrix}
\end{equation}  
These rotations serve as fundamental operators that enable precise control over configuration state mixing.
Importantly, they can preserve critical physical symmetries, notably the particle number conservation and the $z$-projection of the system's total spin, $m_s$. The particle number conservation is manifested through the commutation relation $[\hat G, \hat N]=0$, where $\hat N$ is the particle-number operator. 
Conservation of $m_s$ is achieved by restricting rotations to operations between same-spin orbitals.

In the context of quantum circuit implementation, consider the extension of Eq.~\eqref{eq:rotmatrix} to a $d=2$ Hilbert subspace: 
\begin{widetext}
	\begin{equation}
		\begin{aligned}
			G(\theta)& =\begin{pmatrix}1&0&0&0\\0&\cos(\theta/2)&-\sin(\theta/2)&0\\0&\sin(\theta/2)&\cos(\theta/2)&0\\0&0&0&1\end{pmatrix} :=
			\begin{quantikz} 
				& &\targ{}&\gate{R_Y({\theta}/{2})}&\targ{}&&\\
				& \gate{H}&\ctrl{-1}&\gate{R_Y( {\theta}/{2})}&\ctrl{-1}&\gate{H}&\\ 
			\end{quantikz}.
		\end{aligned}
		\label{eq6}
	\end{equation} 
\end{widetext} 
The rotation $G(\theta)$ can mix the $\ket{01}$ and $\ket{10}$ basis states while leaving the $\ket{00}$ and $\ket{11}$ states unchanged.  Under the JW mapping, this operation corresponds to a single-electron excitation among spin-orbitals (without considering its parity). 

Beyond single excitations, the Givens framework can be extended to describe higher-order excitations. 
In particular, the four-qubit \enquote{double-excitation} Givens rotation  $G^{(2)}(\theta)$, which rotates the $\ket{0011}$ and $\ket{1100}$ states as follows: \begin{equation}\begin{aligned}
		G^{(2)}(\theta) |0011\rangle&=\cos(\theta/2) |0011\rangle+\sin(\theta/2) |1100\rangle ,\\G^{(2)}(\theta) |1100\rangle&=\cos(\theta/2) |1100\rangle-\sin(\theta/2) |0011\rangle .
	\end{aligned} 
\end{equation} 
The decomposition of $G^{(2)}(\theta)$ into one- and two-qubit gates is shown in the Appendix~\ref{ref:appendix}.
In addition, controlled Givens rotations are introduced to guarantee universality in our circuit design. 
Formally, a controlled single-excitation rotation, $CG(\theta)$, acts on a three-qubit subspace by applying a Givens rotation to two target qubits conditioned on the control qubit being in the $\ket{1}$ state: 
\begin{equation}\begin{aligned}
		CG(\theta)\left|1ab\right\rangle&=\left|1\right\rangle\otimes G(\theta)\left|ab\right\rangle,\\ CG(\theta)\left|0ab\right\rangle&=\left|0\right\rangle\otimes\left|ab\right\rangle,
	\end{aligned} 
\end{equation}
where $\ket{ab}$ denotes the state of the two target qubits. Controlled Givens rotations enable the selective excitation of a specific configuration within a superposition. 
As a result, individual components of a multireference state can be independently addressed and coherently manipulated.

\subsection{Givens-based multireference state preparation}
In the JW mapping, we consider the state $\ket{0011}\equiv 
\begin{quantikz}[row sep=0.2cm, column sep=0.2cm]
	& &&\\
	&\ctrl{0}& \octrl{0}&
\end{quantikz}$ 
as a single-reference state with two electrons and four spin-orbitals. Here, 
$\begin{quantikz}[row sep=0.02cm, column sep=0.02cm]\ctrl{0}\end{quantikz}$ 
and 
$\begin{quantikz}[row sep=0.02cm, column sep=0.02cm]\octrl{0}\end{quantikz}$ 
represent the $\alpha$ and $\beta$ electron spins, respectively. This state can, for example, represent the HF state of the $\mathrm{H_2}$ molecule near the equilibrium bond length using a minimal basis set. 

Our approach prepares the multireference state by generating excited configurations from the single-reference state. This is achieved by rotating the diagonal-pair two-qubit $\ket{01}$ and $\ket{10}$ subspaces in the Fock space using Givens rotations. As shown in Fig.~\ref{fig1}~(b), the first Givens rotation $G(\theta_1)$ and the second Givens rotation $G(\theta_2)$ rotate the subspaces of qubits 1 and 3, and qubits 2 and 4, respectively. These rotations are applied to spin-orbitals with the same spin, thereby preserving $m_s$ of the molecular system. The four SDs shown in Fig.~\ref{fig1}~(b) account for all possible spin-conserving configurations in our case:
\begin{equation} 
	a  \begin{quantikz}[row sep=0.2cm, column sep=0.2cm]
		& &&\\
		&\ctrl{0}& \octrl{0}&
	\end{quantikz}
	+ b \begin{quantikz}[row sep=0.2cm, column sep=0.2cm]
		&\ctrl{0} &&\\
		&& \octrl{0}&  
	\end{quantikz}
	+ c \begin{quantikz}[row sep=0.2cm, column sep=0.2cm]
		& &\octrl{0}&\\
		&\ctrl{0}& &  
	\end{quantikz} 
	+d \begin{quantikz}[row sep=0.2cm, column sep=0.2cm]
		&\ctrl{0} &\octrl{0}&\\
		&& &  
	\end{quantikz} ,	
\end{equation} 
where $a$, $b$, $c$, $d$ are the coefficients of the respective SD.

Considering a more complicated example, the ground state of stretched $\mathrm{H_2O~(4e,4o)}$ (with four active electrons and orbitals), when truncated to the leading three configurations, is typically represented as $a\ket{00001111}+b\ket{00110011}+c\ket{11001100}$ (assuming HF molecular orbitals) in the JW mapping. 
This state can be prepared through the sequential application of generalized Givens rotations: $CG^{(2)}_{3; 8,7,2,1}(\theta_2)~G^{(2)}_{6,5,4,3}(\theta_1)\ket{00001111}$. Here, the first subscript in the $CG^{(2)}$ gate indicates the control qubit, while subsequent subscripts denote the target qubits. 
Both $G^{(2)}$ and $CG^{(2)}$ gates exclusively excite the HF determinant $\ket{00001111}$. 

To reduce computational complexity, we adopt the qubit tapering approach that exploits inherent Hamiltonian symmetries, specifically $\mathbb{Z}_2$ Pauli symmetries \cite{57}. This technique maps qubit operators into the optimal symmetry eigensectors, effectively eliminating redundant degrees of freedom and reducing the total qubit count required for the quantum simulation. For our $\mathrm{H_2O~(4e,4o)}$ example, qubit tapering reduces the reference state to a more compact form: $a\ket{00001}+b\ket{00110}+c\ket{11001}$. 
This tapered form is derived by examining the qubits removed by tapering and mapping the corresponding configurations from the full Hilbert space onto the reduced subspace. 

However, in doing so, the transformation disrupts the direct correspondence between spin-orbital occupations and qubits, which means Givens rotations alone are no longer sufficient to construct general MR states. 
To address this, we introduce controlled-$X$ gates, which enables the preparation of specific configurations in the reduced qubit space by conditionally modifying bitstring patterns. 
This provides a mechanism for realizing SDs that cannot be directly generated through Givens rotations alone due to the altered qubit mapping.

For our H$_2$O (4e,4o) example, the final gate sequence implementing the tapered MR state employs a subset of the general gate set $\{CG,G,CX\}$, and consists of the following operations: 
\begin{equation}
	CX_{4,1}~CX_{4,5}~G_{4,1}(\theta_2)~CX_{2,3}~G_{2,1}(\theta_1)\ket{00001},
\end{equation}
where the subscripts in $CX_{i,j}$ denote that $i$ is the control qubit and $j$ is the target qubit. This circuit is manually constructed to reproduce the tapered MR state, with $CX$ gates ensuring correct excitation structure in the reduced qubit space.

The rotation parameters $\theta_1$ and $\theta_2$ are determined through the solution of the following system of equations:
\begin{equation}
	\left\{\begin{array}{l}
		a=\cos \left(\theta_1 / 2\right) \cos \left(\theta_2 / 2\right), \\
		b=\sin \left(\theta_1 / 2\right), \\
		c=\cos \left(\theta_1 / 2\right) \sin \left(\theta_2 / 2\right),
	\end{array}\right. 
\end{equation}
where $a$, $b$, and $c$ again directly correspond to the coefficients of the the target MR states.

\section{Computational Details \label{section3}}

\subsection{Benchmark systems and reference states}
We evaluate the performance improvement of MREM over single-reference REM in treating electron correlation by conducting noisy VQE simulations on three representative small molecular systems: $\mathrm{H_2O}$, $\mathrm{N_2}$, and $\mathrm{F_2}$. 
While the ground states of $\mathrm{N_2}$, $\mathrm{H_2O}$ and $\mathrm{F_2}$ are reasonably well-described by single-reference methods at equilibrium geometry, bond dissociation processes demand methods capable of capturing strong correlation effects. 
These systems form a small, targeted subset for probing strong correlation in small molecules, particularly in bond-breaking regimes, thereby serving as useful test cases for evaluating quantum error mitigation strategies.

Second-quantized Hamiltonians with restricted HF orbitals were computed using \texttt{PySCF} \cite{62}, with correlation-consistent basis sets: cc-pVDZ for $\mathrm{H_2O}$ and $\mathrm{F_2}$, and cc-pVTZ for $\mathrm{N_2}$. 
Active space electronic Hamiltonians were extracted via \texttt{ActiveSpaceTransformer} in \texttt{Qiskit} \cite{63}, with active spaces of $\mathrm{H_2O~(4e, 4o)}$, $\mathrm{N_2~(6e, 6o)}$, and $\mathrm{F_2~(10e, 6o)}$. 
All fermionic Hamiltonians were mapped to qubits using the JW mapping. 
Qubit tapering reduced the number of qubits as follows: $\mathrm{H_2O~(8 \rightarrow 5)}$, $\mathrm{N_2~(12 \rightarrow 8)}$, and $\mathrm{F_2~(12 \rightarrow 8)}$. 
Exact diagonalization of the resulting qubit Hamiltonians was conducted using the \texttt{NumPyMinimumEigensolver} algorithm.

The target MR states were obtained through conventional quantum chemistry methods including CISD, CCSD (using \texttt{PySCF}), and DMRG (using \texttt{block2} \cite{64}). The resulting wavefunctions were converted into state vectors using the \texttt{import\_state} function in \texttt{PennyLane} \cite{65}, from which the relevant Slater determinants and their coefficients were extracted \cite{77}. 

Approximate MR states were constructed by selecting 2–3 dominant Slater determinants based on their weight in the full wavefunction. For $\mathrm{H_2O}$, we used CCSD with a 6-31G basis set; for $\mathrm{F_2}$, CISD with STO-6G; and for $\mathrm{N_2}$, DMRG with cc-pVTZ. 
We note that the basis set used to generate the MR state does not need to match that of the target Hamiltonian; 
smaller basis sets can still provide sufficiently accurate coefficients for the limited number of retained determinants, significantly reducing classical computational cost.

For the wavefunction ansatz, we employed a hardware-efficient ansatz (HEA), specifically the $R_Y$-linear ansatz $\hat U_{R_Y}(\boldsymbol{\theta})$, with 5 layers for H$_2$O, F$_2$, and 20 layers for N$_2$ \cite{54}. 
This ansatz is well suited to near-term quantum hardware due to its shallow depth, compatibility with native gate sets, and adaptability to device-specific connectivity constraints.
However, it is known to suffer from trainability and optimization challenges \cite{McClean2018}, particularly in systems involving more than 20 qubits.

In the context of MREM, the interplay between the ansatz and state preparation circuits also requires careful consideration, especially regarding their ordering and initialization behavior on quantum hardware.
Concerning the initial state preparation, it is important to point out that $U_{R_Y}$ only acts trivially as the identity on the all-0 state, $U_{R_Y}(0) \ket{0} = \ket{0}$ for $\boldsymbol{\theta} = 0$. 
Finding parameters for which $U_{R_Y}(\boldsymbol{\theta})$ acts as the identity on a general initial state is not trivial. 
Thus, we place the $R_Y$-linear ansätze before the state preparation circuit, \textit{i.e.}, $\hat U_{\mathrm{init}} \hat U_{R_Y}\ket{0}$ instead of $\hat U_{R_Y} \hat U_{\mathrm{init}}\ket{0} $, as shown in Fig.~\ref{fig1}(c).	

Table~\ref{tab1} summarizes key variables for each system, including the number of qubits after tapering, ansatz depth, reference wavefunction source, and the number of Slater determinants used.

\begin{table}
	\centering
	\caption{The number of qubits (\# qbs), the number of repeated layers, $L$, in the $R_Y$-linear ansätze, the theoretical sources for MR states, and the number of prepared SDs for each simulated system.}
	\begin{tabular}{lcclc}
		\hline \hline
		System  & \# qbs & $L$ & MR source & \# SDs  \\ \hline
		H$_2$O (4e, 4o; cc-pVDZ) & 5 & 5 & CCSD/6-31G & 3 \\
		F$_2$ (10e, 6o; cc-pVDZ)  & 8 & 5 & CISD/STO-6G & 2 \\
		N$_2$ (6e, 6o; cc-pVTZ)  & 8 & 20 & DMRG/cc-pVTZ & 3 \\
		\hline \hline
	\end{tabular}
	\label{tab1}
\end{table}

\subsection{Quantum circuit implementation and noise modeling}
Quantum circuits and the VQE algorithm were implemented using the \texttt{Estimator} module of \texttt{Qiskit Aer 0.13.1}. 
For all \texttt{Estimator} simulations, each energy evaluation was estimated using $1 \times 10^7$ sampling shots. 
To accelerate simulations, we enabled the \texttt{approximation} option of the \texttt{Estimator}, which approximates the sampling distribution of measurement outcomes as a normal distribution. This \texttt{approximation} method significantly improves efficiency by avoiding explicit sampling, while still incorporating statistical noise. However, this method does not model readout errors, and therefore our results primarily reflect the effects of gate noise.	

Noise simulations were performed using the \texttt{FakeSydneyV2} backend noise model, which incorporates depolarization and thermal relaxation errors on all single- and two-qubit gates. Device-specific noise parameters, including gate errors, durations, readout errors, and decoherence times, were derived from real IBM device calibration data \cite{66}.

For variational optimization, we used a gradient-free classical optimizer based on the implicit filtering (ImFil) algorithm, as implemented in the \texttt{scikit-quant} package \cite{67}. This optimizer is well suited for noisy, high-dimensional landscapes with many local minima.

\subsection{Enforcing Spin Symmetry}
HEA, while favored for their low circuit depth and hardware compatibility, do not inherently preserve physical symmetries.
This lack of symmetry adaptation can result in qualitatively incorrect variational states, manifesting as nondifferentiable cusps in potential energy surfaces (PES) and significant spin contamination \cite{68}.
Such issues are particularly pronounced in systems like $\mathrm{H_2O}$ and $\mathrm{N_2}$, where an accurate description of the singlet ground state is essential.
Quantum noise further exacerbates symmetry breaking, as gate errors and decoherence can push the variational state out of the desired symmetry sector.

To address these symmetry-breaking effects, we introduce a spin penalty term into the qubit Hamiltonian: 
$\hat{H}^\prime = \hat{H} + \lambda \cdot \hat{\mathbf{S}}^2,$  
where $\hat{\mathbf{S}}^2$ is the total spin angular momentum operator and $\lambda > 0$ is a tunable penalty coefficient. 
Since our goal is to recover the singlet ground state ($S = 0$), this penalty lowers the energy of spin-pure singlet states relative to spin-contaminated alternatives. 
Importantly, this linear penalty form exploits the fact that the eigenvalues of $\hat{\mathbf{S}}^2$ are non-negative and minimized for singlets.  
Unlike the standard squared deviation formulation, $\lambda \cdot (\hat{\mathbf{S}}^2 - S(S+1))^2$ \cite{28}, this linear variant avoids introducing a large number of additional Pauli terms, thereby reducing measurement overhead in noisy simulations.

In our simulations, we set $\lambda = 0.1$ for H$_2$O and $\lambda = 0.5$ for N$_2$, based on empirical tuning.
These values were sufficient to suppress spin contamination and stabilize VQE convergence without significantly distorting the underlying energy landscape.
For F$_2$, where the HEA did not lead to appreciable symmetry breaking, no penalty term was applied.

\section{Results and Discussion \label{section4}}
\begin{figure*}
	\centering    \includegraphics[width=\linewidth]{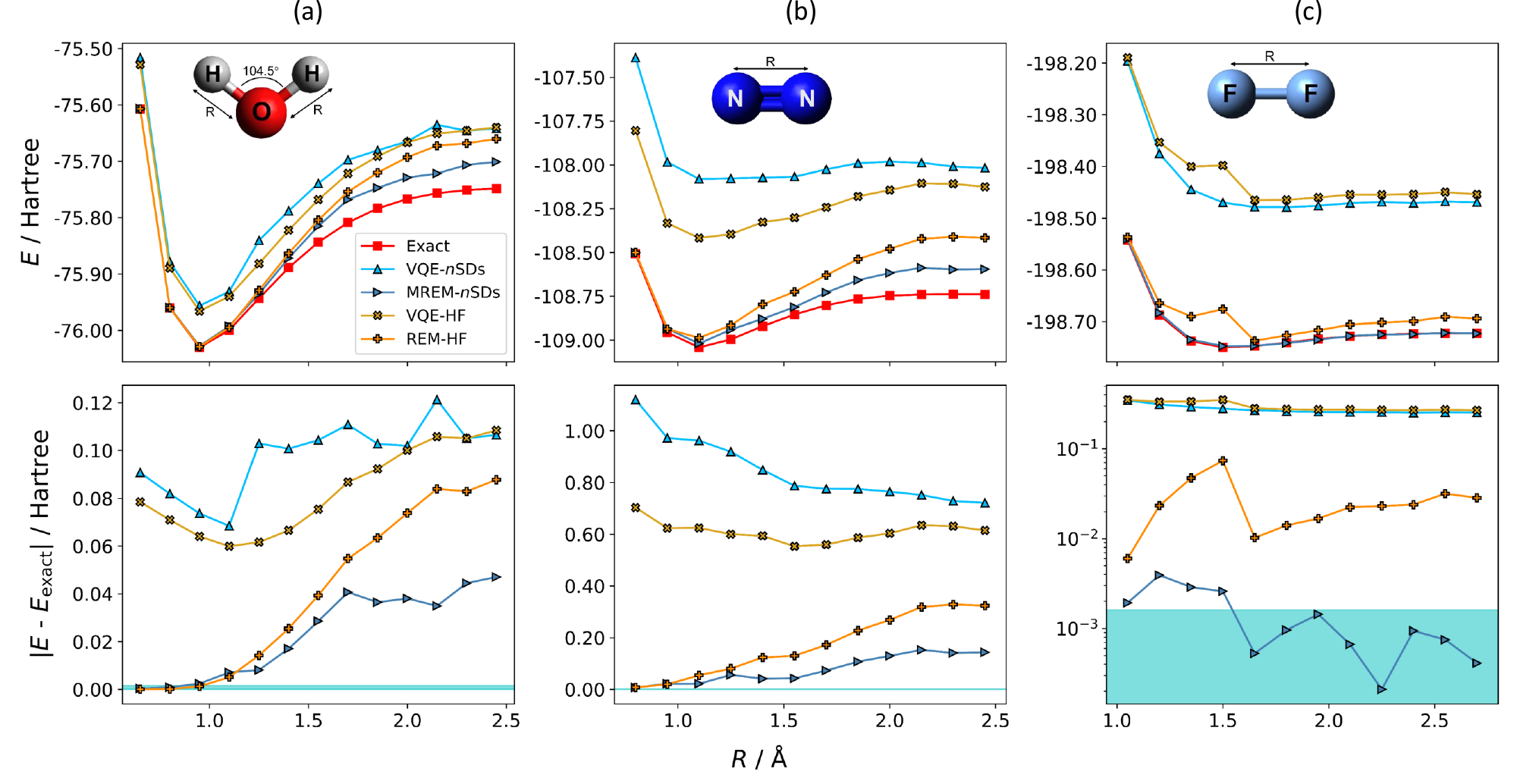}
	\caption{Comparisons of MREM and single-reference REM for PESs. PESs (top) and absolute errors (bottom) are computed for (a) $\mathrm{H_2O}$ (4e, 4o; cc-pVDZ) symmetric stretching, (b) $\mathrm{N_2}$ (6e, 6o; cc-pVTZ), and (c) $\mathrm{F_2}$ (10e, 6o; cc-pVDZ) bond dissociation. The cyan-shaded areas represent computational accuracy below $1.6 \times 10^{-3}$ Hartree (1 kcal/mol).}
	\label{fig2}
\end{figure*}

In this section, we evaluate the performance of MREM by computing potential energy surfaces for our collection of molecules: $\mathrm{H_2O}$, $\mathrm{N_2}$, and $\mathrm{F_2}$. Fig.~\ref{fig2} shows comparisons between MREM and single-reference REM. In this figure, VQE results using an HF initial state is labeled as \enquote{VQE-HF}, while the corresponding REM-corrected curve is denoted by \enquote{REM-HF}.
Similarly, the VQE data calculated using a linear combination of $n$ SDs (Table~\ref{tab1}) as initial state (and reference) is labeled \enquote{VQE-$n$SDs}, while its corresponding MREM-correction is denoted by \enquote{MREM-$n$SDs}. 

Note also that in these tests, we compare to a \textit{computational accuracy}. The latter threshold is defined as an error of $1.6 \times 10^{-3}$ Hartree (1 kcal/mol) with respect to the exact result obtained in the complete absence of noise, using the same level of theory. We make this point because a given level of theory need not be exact with respect to reality. In other words, a calculation with computational accuracy need not have the \textit{chemical accuracy}  needed for realistic chemical predictions, a distinction suggested in~\cite{22}.	

For $\mathrm{H_2O}$ the MR state is constructed using three dominant Slater determinants via two-qubit Givens rotations (Eq.~\eqref{eq6}), enabling efficient generation of single excitations. 
Due to the additional circuit complexity, the VQE-3SDs results exhibit higher energy compared to VQE-HF, as depicted in Fig.~\ref{fig2}a. 
However, after applying the REM correction, MREM-3SDs yields a substantial reduction in error and recovers a more accurate potential energy surface. 
This highlights the importance of combining MR states with error mitigation: although MR states alone may not improve noisy VQE outcomes, they serve as a more expressive and physically grounded reference for the mitigation step. 
The gate-efficient MR construction thus enables MREM to extract meaningful physical information in noisy conditions, while keeping circuit overhead moderate.

The $\mathrm{N_2}$ molecule (Fig.~\ref{fig2}b) presents a greater challenge due to its strong multireference character. 
Capturing the relevant correlation requires MR states with double excitations, implemented using four-qubit Givens rotations ($G^{(2)}$). 
However, the decomposition of $G^{(2)}$ is not gate-efficient (see the Appendix~\ref{ref:appendix}), resulting in substantial circuit depth and elevated noise. 
Consequently, the unmitigated VQE-3SDs performs worse than VQE-HF in both energy accuracy and the overall PES profile. 
Nonetheless, the error-mitigated MREM-3SDs recovers much of the expected physical behavior in the bond-stretching region. 
The improved PES shape and consistently reduced errors demonstrate that when facing a balance between expressivity and circuit complexity, using MR states can provide clear advantages when combined with error mitigation.

The F$_2$ molecule (Fig.~\ref{fig2}c) is a prototypical example where a compact MR state with two selected SDs suffices to capture the essential near-degeneracy between bonding $\sigma$ and antibonding $\sigma^\star$ orbitals during bond-stretching region. 
This leads to near-computational accuracy in MREM-2SDs, with errors reduced by approximately two orders of magnitude compared to the noisy VQE results and by one order of magnitude compared to the REM-HF results. 
Despite the additional noise in MREM-2SDs circuits, the MR state provides a better initialization that improves convergence during VQE optimization, particularly in regions where $R \leq\mathrm{1.5~\AA}$, where noisy VQE often fails due to local minima.   

\section{Conclusion and Outlook \label{section5}}
In this work, we have developed a multireference error mitigation (MREM) framework for improving quantum simulations of strongly correlated molecular systems. Central to this approach is the preparation of multireference (MR) states that exhibit substantial overlap with the target ground state, constructed from conventional quantum chemistry methods such as configuration interaction, coupled cluster, or the density matrix renormalization group.

We introduce an efficient and physically motivated scheme to prepare these states on quantum hardware using Givens rotations, and validate the resulting Givens-based MREM framework through noisy digital quantum simulations of H$_2$O, N$_2$, and F$_2$. 
Our results demonstrate two key advantages: (i) MR references enhance REM performance over single-determinant schemes, providing physically motivated energy  error mitigation at low cost, and (ii) Givens-encoded MR states serve as robust initializations for VQE, reducing the risk of becoming trapped in local minima and accelerating convergence (see the Appendix~\ref{ref:appendix}).
While increasing the number of reference states can introduce additional circuit noise, our results suggest that the improved mitigation performance outweighs this cost -- particularly in shallow circuit regimes relevant to near-term hardware. 

MREM is an error mitigation framework that can be integrated with a broad class of variational quantum algorithms beyond VQE, offering flexibility for near-term quantum chemistry applications. 
Within this framework, the primary scalability challenge arises from the circuit complexity required for MR state preparation, which represents the most essential and resource-intensive subroutine.
To address this resource bottleneck, future work will focus on developing more compact circuit constructions that preserve essential physical symmetries while reducing gate overhead. 
Promising directions include combining MREM with spin-conserving methods \cite{40, Berry2024, Moerchen2024, Ollitrault2024, Burton2023, Sugisaki2019, Sugisaki2016, Carbone2022, Tew2024, Streif2024, Dobrautz2019, LiManni2020, LiManni2021, Dobrautz2022}, explicitly correlated methods \cite{Kong2011, Grneis2017, Httig2011, Schleich2022, Volkmann2024}, especially the transcorrelated approach \cite{Boys1969, Cohen_2019, Haupt2023, Dobrautz2022b, Dobrautz2019b, Sokolov2023, Motta2020, McArdle2020, Kumar2022, 55}, tiled unitary product states \cite{Burton2024, 70} as well as the separable-pair approximation \cite{Kottmann2022, Kottmann2023}. 

Another promising avenue for enhancing MREM lies in exploring the expressivity differences between Clifford and near-Clifford ansatz states. 
For instance, an open question is whether Clifford circuit initialization, which restricts single-qubit rotation gates in the HEA to discrete multiples of $\pi/2$ \cite{81}, can enhance the performance of our Givens-based MR circuit in approximating the ground state. 

We have also identified several broader opportunities for enhancing the MREM framework itself. 
Because MREM achieves noise reduction by exploiting the inherent classical simulatability of select MR circuits, one can imagine viewing such circuits as being virtual post-processing operators expressed as $\underbrace{U_{\mathrm{init}}}_{\mathrm{class.}}\underbrace{U_{R_Y}}_{\mathrm{quant.}}\ket{0}$, in accordance with the Schrödinger-Heisenberg VQE paradigm \cite{78}. 
Such an approach could strategically offload circuit complexity to classical devices, leading to shallower quantum circuits that are more resilient to noise.  
Finally, given MREM's fixed circuit structure designed to prevent noisy gate variable effects, its integration with adaptive ansätze \cite{grimsley_adaptive_2019, Magnusson2024} using statistical tools merits investigation.

\section{Supporting Information}

Supporting Information (SI) is available for this article, see Sec.~\ref{ref:appendix}.
The SI contains information on the decomposition of $G^{(2)}(\theta)$ into one- and two-qubit gates; details (coefficients and configurations) of the MR states for each simulated system and distance; gate resource statistics of the $R_Y$-linear and MR state preparation circuits of all studied systems; numerical values of the results shown in Fig.~\ref{fig2}; and details on improved VQE convergence behavior due to MR initial states. 

\section{Acknowledgments}

Funded by the European Union. 
Views and opinions expressed are, however, those of the author(s) only and do not necessarily reflect those of the European Union or REA. 
Neither the European Union nor the granting authority can be held responsible for them.
This work was funded by the EU Flagship on Quantum Technology HORIZON-CL4-2022-QUANTUM-01-SGA project 101113946 OpenSuperQPlus100. 
This research has been supported by funding from the Wallenberg Center for Quantum Technology (WACQT).
WD acknowledges funding from the European Union’s Horizon Europe research and innovation program under the Marie Sk{\l}odowska-Curie grant agreement No. 101062864 and the German Federal Ministry of Education and Research (BMBF) under the research program \textit{Quantensysteme} and funding measure \textit{Quantum Futur 3} for project No. 13N17229. 

This research relied on computational resources provided by the National Academic Infrastructure for Supercomputing in Sweden (NAISS) at C3SE and NSC, partially funded by the Swedish Research Council through grant agreement no. 2022-06725.

\section*{\label{ref:appendix} Appendix}
\appendix


\section{Gate decomposition \label{g2gate}}
\begin{figure*}
	\resizebox{1\linewidth}{!}{ \begin{quantikz} 
			& \gate[4]{G^{(2)}(\theta)}& \\
			&&\\&&\\&&\\
		\end{quantikz} :=
		\begin{quantikz} 
			&  &  \ctrl{2} & \gate{H} & \ctrl{1} & \gate{R_Y(-\frac{\theta}{8})} & \ctrl{3} & \qw &  & \gate{R_Y(-\frac{\theta}{8})} &  & \targ{} & \gate{R_Y(\frac{\theta}{8})} & \qw & & \ctrl{3} & \gate{R_Y(\frac{\theta}{8})} & \ctrl{1} & \targ{} & \gate{H} & \ctrl{2} & \qw &\\
			&  &  \qw & \qw &  \targ{}  & \gate{R_Y(\frac{\theta}{8})}  &   & \qw & \targ{} & \gate{R_Y(\frac{\theta}{8})} &\targ{}& \qw & \gate{R_Y(-\frac{\theta}{8})} &\targ{} &   &  & \gate{R_Y(-\frac{\theta}{8})}& \targ{}& \qw & \qw & \qw & \qw &\\
			& \ctrl{1} & \targ{} & \qw &  \ctrl{1} & \qw &  & \qw &   &  & \ctrl{-1}& \ctrl{-2} & \qw & \qw &   & \qw & \qw & \qw & \ctrl{-2} & & \targ{}&\ctrl{1} & \qw \\
			& \targ{} & \qw & \gate{H} & \targ{} & \qw & \targ{} &  \gate{H}&\ctrl{-2} &   & \qw & \qw & \qw & \ctrl{-2}& \gate{H} &\targ{} & \qw & \qw & \qw &\gate{H} &&\targ{}&  \\
		\end{quantikz}.
	} 
	\caption{\label{fig:g2-decomp}Decomposition of $G^{(2)}(\theta)$ into Hadamard, $R_Y$ rotational gates, and CNOT gates.}
\end{figure*}
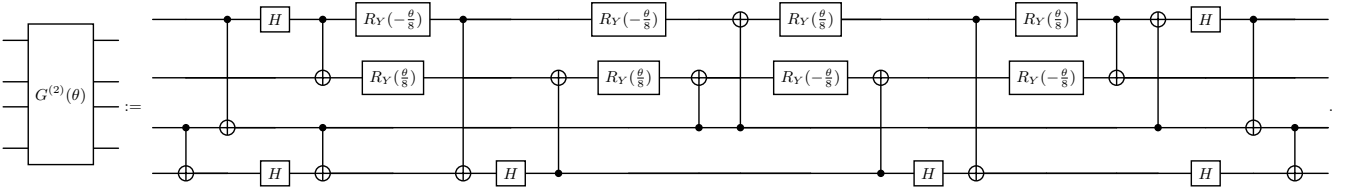

The decomposition of $G^{(2)}(\theta)$ into single-qubit Hadamard and $R_Y$ rotational gates as well as two-qubit CNOT gates is shown in Fig.~\ref{fig:g2-decomp}.  
Due to the relatively high error rates of two-qubit gates, we adopt an optimal scheme with fewer two-qubit gates. 
$G(\theta)$ uses 2 CNOT gates (see Eq.~(6) of the main text), while $G^{(2)}(\theta)$ uses 14 CNOT gates.	

\section{Multireference states and circuits details}
\begin{figure*}
	\centering
	\begin{minipage}{0.45\textwidth}
		\centering
		\begin{quantikz}[transparent,row sep=0.2cm]
			\put(-20,5){\textbf{(a)}}
			&\gate{X} &\gate[2]{G(\theta_1)} & &\gate{G(\theta_2)}  &\gate{X}&\\
			&& & \ctrl{1}&\wire[u]{q} &\wire[u]{q}&\\
			&& & \gate{X}&\wire[u]{q}&\wire[u]{q}&\\
			&&&&\gate{G(\theta_2)}\wire[u]{q}&\ctrl{1}\wire[u]{q}&\\
			&&&&&\gate{X}\wire[u]{q}&
		\end{quantikz}
	\end{minipage}
	\begin{minipage}{0.45\textwidth}
		\centering
		\begin{quantikz}[transparent,row sep=0.2cm]
			\put(-20,5){\textbf{(b)}}
			&\gate{X}&\gate{G(\theta_1)}& &\gate[2]{G(\theta_2)} &\gate{X}&\\
			&&\wire[u]{q} & &&\ctrl{1}\wire[u]{q}&\\
			& &\wire[u]{q}& &&\gate{X}\wire[u]{q}&\\
			&& \gate{G(\theta_1)}\wire[u]{q}&\ctrl{1}&& &\\
			&&&\gate{X}\wire[u]{q}&& &
		\end{quantikz}
	\end{minipage}
	
	\vspace{0.5cm}
	
	\begin{minipage}{0.3\textwidth}
		\centering
		\begin{quantikz}[transparent,row sep=0.2cm]
			\put(-20,5){\textbf{(c)}}
			&\gate{X}& & &\\
			&\gate{X}& & &\\
			&\gate{X}& \gate{G(\theta_1)}& & \\
			&\gate{X}&\wire[u]{q} &&\\
			&\gate{X}&\wire[u]{q} &&\\
			&\gate{X}&\wire[u]{q} &&\\
			&&\gate{G(\theta_1)}\wire[u]{q}&\ctrl{1}&\\
			&&&\gate{X}&
		\end{quantikz}
	\end{minipage}
	\begin{minipage}{0.3\textwidth}
		\centering
		\begin{quantikz}[transparent,row sep=0.2cm]
			\put(-20,5){\textbf{(d)}}
			&\gate{X}& & &\\
			&\gate{X}& & &\\
			&\gate{X}& & &\\
			&\gate{X}&&&\\
			&\gate{X}&&&\\
			&\gate{X}&\gate[2]{G(\theta_1)}&&\\
			&&&\ctrl{1}&\\
			&&&\gate{X}&
		\end{quantikz}
	\end{minipage}
	\begin{minipage}{0.3\textwidth}
		\centering
		\begin{quantikz}[transparent,row sep=0.2cm]
			\put(-20,5){\textbf{(e)}}
			&\gate{X}& & &\\
			&\gate{X}&\gate{G^{(2)}(\theta_1)} &\gate{G(\theta_2)} &\\
			&\gate{X}& \gate{G^{(2)}(\theta_1)} \wire[u]{q}&\ctrl{1}\wire[u]{q}& \\ &&\wire[u]{q}&\gate{G(\theta_2)}&\\
			&&\gate{G^{(2)}(\theta_1)} \wire[u]{q}&&\\
			&\qw&\gate{G^{(2)}(\theta_1)}\wire[u]{q}&&\\
			&\qw&&&\\
			\\
			&\qw&&&
		\end{quantikz}
	\end{minipage} 
	
	\caption{Specific circuits for the MR states preparation.  (a) H$_2$O, $R:$ all calculations except for $R=2.15$ \AA; (b) H$_2$O,  $R = 2.15$ \AA; (c) F$_2$, $R: 1.05$ \AA $- 1.5$ \AA ; (d) F$_2$, $R: 1.65$ \AA $ - 2.70$ \AA;  (e) N$_2$, $R:$ all calculations.}
	\label{all_MR_circuits}
\end{figure*}
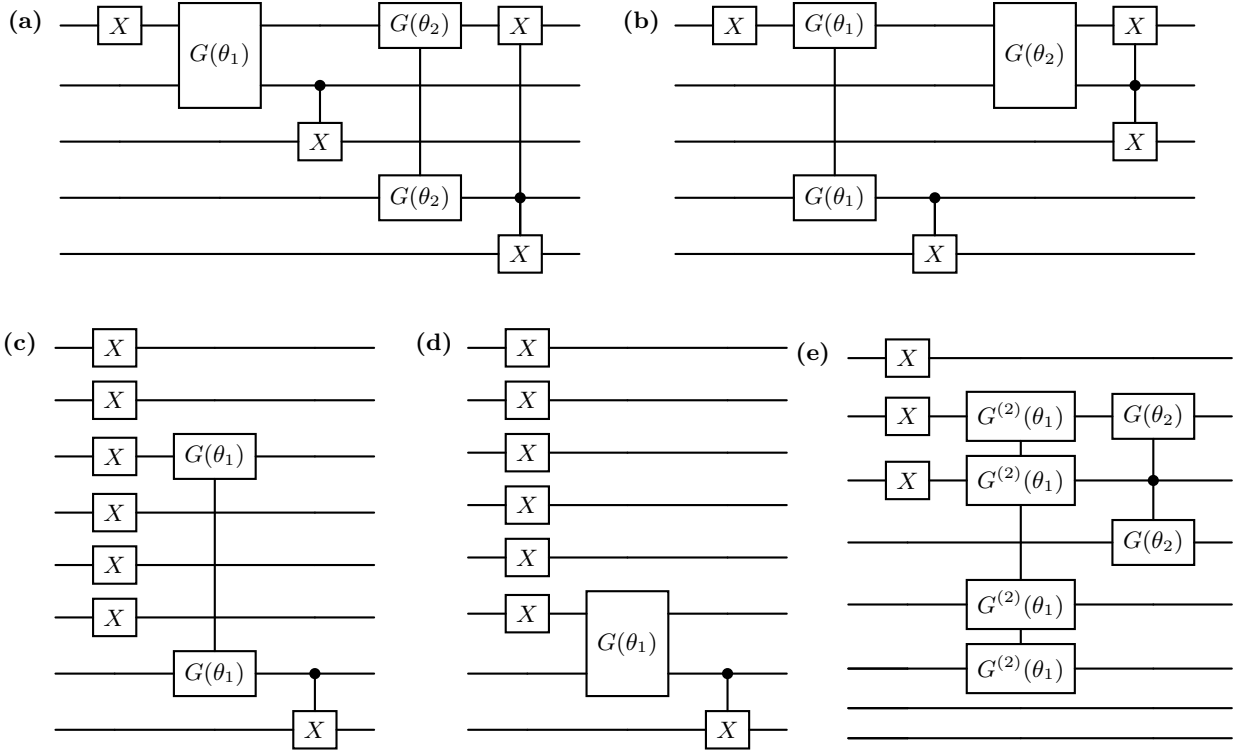

Table~\ref{chosen_mr_states} details the prepared MR states for each simulated system, and Fig.~\ref{all_MR_circuits} illustrates the corresponding MR state preparation circuits. For cases where the choices coincide with HF states, the classical CCSD and DMRG solvers provide solutions that closely approximate the HF states (in the active spaces). Therefore, we retain the MR state preparation circuits with parameters initialized to zero, ensuring the consistency of noisy gates. Furthermore, classical calculations indicate that there are distinctly different leading excited configurations for $\mathrm{H_2O}$ at $R = 2.15$~\AA. 

\begin{table*}[h]
	\centering
	\caption{Selected MR states and preparation parameters for $\mathrm{H_2O}$, $\mathrm{F_2}$, and $\mathrm{N_2}$ based on the circuits from Fig.~\ref{all_MR_circuits}. The MR state preparations for $\mathrm{H_2O}$ and $\mathrm{N_2}$ require two parameters, labeled as ($\theta_1$, $\theta_2$) in Fig.~\ref{all_MR_circuits}.
	}
	\label{chosen_mr_states}
	\begin{tabular}{ccccc}
		\hline\hline
		Molecule & $R$ / \AA & MR state & Parameters \\ \hline
		\multirow{13}{*}{H$_2$O} &0.65 & $\ket{00001} - 0 \cdot \ket{11001} - 0 \cdot\ket{00110}$ & (0, 0)\\  
		&0.80 & $\ket{00001} - 0  \cdot\ket{11001} - 0 \cdot \ket{00110}$ & (0, 0)\\ 
		&0.95 & $\ket{00001} - 0 \cdot \ket{11001} - 0 \cdot \ket{00110}$ & (0, 0) \\ 
		&1.10 & $\ket{00001} - 0 \cdot \ket{11001} - 0 \cdot \ket{00110}$ & (0, 0) \\ 
		&1.25 & $0.9929 \ket{00001} - 0.0906 \ket{11001} - 0.0768 \ket{00110}$ &(-0.1543, -0.1815)\\ 
		&1.40 & $0.9852 \ket{00001} - 0.1218 \ket{11001} - 0.1204 \ket{00110}$ &(-0.2443, -0.2432)\\ 
		&1.55 & $0.9726 \ket{00001} - 0.1708 \ket{00110} - 0.1574 \ket{11001}$ &(-0.3434, -0.3209)\\ 
		&1.70 & $0.9535 \ket{00001} - 0.2265 \ket{00110} - 0.1990 \ket{11001}$ &(-0.4570, -0.4116)\\ 
		&1.85 & $0.9254 \ket{00001} - 0.2867 \ket{00110} - 0.2480 \ket{11001}$ &(-0.5815, -0.5237)\\ 
		&2.00 & $0.8867 \ket{00001} - 0.3490 \ket{00110} - 0.3031 \ket{11001}$ & (-0.7131, -0.6587)\\ 
		&2.15 & $0.8402 \ket{00001} - 0.4048 \ket{00111} - 0.3607 \ket{11000}$ &
		(-0.7381, -0.8980)\\ 
		&2.30 & $0.7933 \ket{00001} - 0.4508 \ket{00110} - 0.4093 \ket{11001}$ & (-0.9353, -0.9526)\\ 
		&2.45 & $0.7529 \ket{00001} - 0.4827 \ket{00110} - 0.4473 \ket{11001}$ & (-1.0075, -1.0720)\\ 
		\hline  
		\hline
		\multirow{12}{*}{F$_2$} &1.05 & $0.9966 \ket{00111111} - 0.0825 \ket{11111011}$ & -0.1652\\ 
		&1.20 & $0.9901 \ket{00111111} - 0.1407 \ket{11111011}$ &-0.2823 \\  
		&1.35 & $0.9761 \ket{00111111} - 0.2171 \ket{11111011}$ & -0.4377\\  
		&1.50 & $0.9517 \ket{00111111} - 0.3072 \ket{11111011}$ & -0.6244 \\  
		&1.65 & $0.9157 \ket{00111111} - 0.4019 \ket{11011111}$ & -0.8271 \\ 
		&1.80 & $0.8723 \ket{00111111} - 0.4891 \ket{11011111}$ &-1.0220\\ 
		&1.95 & $0.8289 \ket{00111111} - 0.5595 \ket{11011111}$ &-1.1875 \\ 
		&2.10 & $0.7918 \ket{00111111} - 0.6107 \ket{11011111}$ &-1.3140 \\  
		&2.25 & $0.7636 \ket{00111111} - 0.6456 \ket{11011111}$ &-1.4037\\ 
		&2.40 & $0.7437 \ket{00111111} - 0.6685 \ket{11011111}$ &-1.4643\\ 
		&2.55 & $0.7303 \ket{00111111} - 0.6831 \ket{11011111}$ &-1.5040\\ 
		&2.70 & $0.7216 \ket{00111111} - 0.6923 \ket{11011111}$ &-1.5294\\ \hline
		\hline
		\multirow{12}{*}{N$_2$} &0.80 & $\ket{00000111} -0 \cdot \ket{00110001} -0 \cdot\ket{00001110}$ & (0, 0) \\ 
		&0.95 & $\ket{00000111} -0 \cdot \ket{00110001} -0 \cdot\ket{00001110}$ & (0, 0)\\ 
		&1.10 & $0.9839 \ket{00000111} -0.1263 \ket{00110001} -0.1263\ket{00001110}$& (-0.2533, -0.2554) \\ 
		&1.25 & $0.9897 \ket{00000111} -0.1012 \ket{00110001} -0.1012\ket{00001110}$& (-0.2027, -0.2038)  \\ 
		&1.40 & $0.9585 \ket{00000111} -0.2016 \ket{00110001} -0.2016\ket{00001110}$ &(-0.4059, -0.4145) \\ 
		&1.55 & $0.9348 \ket{00000111} -0.2512 \ket{00110001} -0.2512\ket{00001110}$ &(-0.5079, -0.5251) \\ 
		&1.70 & $0.8925 \ket{00000111} -0.3189 \ket{00110001} -0.3189\ket{00001110}$ &(-0.6492, -0.6863) \\ 
		&1.85 & $0.8298 \ket{00000111} -0.3946 \ket{00110001} -0.3946\ket{00001110}$ &(-0.8114, -0.8879) \\ 
		&2.00 & $0.7673 \ket{00000111} -0.4534 \ket{00110001} -0.4534\ket{00001110}$ &(-0.9412, -1.0675) \\ 
		&2.15 & $0.7194 \ket{00000111} -0.4911 \ket{00110001} -0.4911\ket{00001110}$ & (-1.0268, -1.1980) \\ 
		&2.30 & $0.6837 \ket{00000111} -0.5160 \ket{00110001} -0.5160\ket{00001110}$ & (-1.0844, -1.2930) \\ 
		&2.45 & $0.6631 \ket{00000111} -0.5293 \ket{00110001} -0.5293\ket{00001110}$ & (-1.1156, -1.3474) \\ \hline\hline
	\end{tabular}
\end{table*}

\section{Gate resources statistics \label{gatenumber}}
Table \ref{tab2} lists the single- and two-qubit gates used in the $R_Y$-linear ansätze and the MR state preparation circuits (following complete decomposition). The prepared MR states for $\mathrm{H_2O}$ and $\mathrm{F_2}$ use the two-qubit Givens rotation $G$, while the target MR states for $\mathrm{N_2}$ use the four-qubit Givens rotation $G^{(2)}$. 
From the noise perspective, the use of $G^{(2)}$ gates and the $N_L=20$ layers (Table.~I of the main text) of the $R_Y$-linear ansätze has a significant detrimental impact on the VQE energy landscape. 
Additionally, even with more layers, the $R_Y$-linear ansatz struggles to achieve computational accuracy in noiseless simulations of the strongly correlated N$_2$ molecule \cite{71}. 
$N_L=20$ is a compromise between the expressibility of the $R_Y$-linear ansatz and the severe negative effect of additional noise on the energy expectation values due to deeper circuits. 
The $R_Y$ ansatz with full connectivity can achieve computational accuracy in noiseless simulations, but the high error rates of numerous two-qubit gates cause the VQE algorithm to fail in noisy simulations. 
Ref.~\cite{71} proposed suggestions for designing hardware-efficient ansätze with greater accuracy and scalability, which show promise in solving the challenging task of strongly correlated molecules like N$_2$.\\ 
\begin{table}[H]
	\centering 
	\caption{Number of single- ($n_1$) and two-qubit gates ($n_2$) in the ${R_Y}$-linear ansätze and the HF/MR preparation circuits for each molecule. 
		The HF state preparation only requires single-qubit gates. 
	}
	\begin{tabular}{ccccccc} \hline\hline
		\multirow{2}{*}{Molecule} & \multicolumn{2}{c}{\underline{$R_Y$ ansatz}}& \underline{HF circuit}  &\multicolumn{2}{c}{\underline{MR circuit}} \\ 
		&  ${n_1}$ & ${n_2}$& $n_1$ &${n_1}$ & ${n_2}$ \\ \hline
		$\mathrm{H_2O}$ & 30 & 20& 1 &9 & 7 \\
		$\mathrm{F_2}$ &48 & 35& 6&10 & 3 \\ 
		$\mathrm{N_2}$ & 168 & 140 & 3 & 57 & 42 \\ 
		\hline\hline
	\end{tabular}
	\label{tab2}
\end{table}

\section{Energy data details}
Numerical values for all data points in Fig.~2 of the main text are listed in Table~\ref{all_energies}.

\begin{table*}[h]
	\centering
	\caption{Summary of numerical results of Fig.~2 of the main text: exact values, noisy VQE energies with HF and MR state preparation circuits, MREM and REM energies, along with (absolute) errors. All energies are in Hartrees.}
	\label{all_energies}
	\begin{tabular}{ccccccccccc}
		\toprule
		\multirow{2}{*}{Molecule} & \multirow{2}{*}{$R$ / \AA} & \multirow{2}{*}{Exact Energy} & \multicolumn{2}{c}{VQE-$n$SDs} & \multicolumn{2}{c}{VQE-HF} & \multicolumn{2}{c}{MREM-$n$SDs} & \multicolumn{2}{c}{REM-HF} \\  \cline{4-11}
		&  &  & Energy & Error & Energy & Error & Energy & Error &  Energy & Error \\ \hline
		
		\multirow{13}{*}{H$_2$O}
		& 0.65 & -75.6069 & -75.5162 & 0.0908 & -75.5284 & 0.0785 & -75.6066 & 0.0003 & -75.6070 & 0.0001 \\ 
		& 0.80 & -75.9605 & -75.8786 & 0.0819 & -75.8895 & 0.0709 & -75.9596 & 0.0009 & -75.9603 & 0.0001 \\ 
		& 0.95 & -76.0302 & -75.9565 & 0.0737 & -75.9661 & 0.0641 & -76.0278 & 0.0024 & -76.0290 & 0.0012 \\ 
		& 1.10 & -75.9998 & -75.9313 & 0.0685 & -75.9399 & 0.0599 & -75.9927 & 0.0071 & -75.9947 & 0.0051 \\ 
		& 1.25 & -75.9434 & -75.8405 & 0.1029 & -75.8818 & 0.0616 & -75.9353 & 0.0081 & -75.9292 & 0.0143 \\ 
		& 1.40 & -75.8889 & -75.7881 & 0.1008 & -75.8223 & 0.0666 & -75.8718 & 0.0171 & -75.8634 & 0.0255 \\ 
		& 1.55 & -75.8436 & -75.7392 & 0.1043 & -75.7681 & 0.0755 & -75.8151 & 0.0285 & -75.8043 & 0.0393 \\ 
		& 1.70 & -75.8087 & -75.6978 & 0.1108 & -75.7218 & 0.0869 & -75.7680 & 0.0407 & -75.7539 & 0.0547 \\ 
		& 1.85 & -75.7836 & -75.6808 & 0.1028 & -75.6913 & 0.0923 & -75.7471 & 0.0364 & -75.7202 & 0.0634 \\ 
		& 2.00 & -75.7669 & -75.6650 & 0.1019 & -75.6668 & 0.1001 & -75.7289 & 0.0380 & -75.6931 & 0.0738 \\ 
		& 2.15 & -75.7566 & -75.6353 & 0.1213 & -75.6509 & 0.1058 & -75.7218 & 0.0349 & -75.6726 & 0.0840 \\ 
		& 2.30 & -75.7511 & -75.6460 & 0.1051 & -75.6461 & 0.1051 & -75.7067 & 0.0444 & -75.6682 & 0.0829 \\ 
		& 2.45 & -75.7483 & -75.6418 & 0.1065 & -75.6399 & 0.1084 & -75.7012 & 0.0471 & -75.6605 & 0.0878 \\ 
		
		\hline
		\multirow{12}{*}{F$_2$} 
		& 1.05 & -198.5425 & -198.1963 & 0.3462 & -198.1897 & 0.3528 & -198.5406 & 0.0019 & -198.5365 & 0.0060 \\ 
		& 1.20 & -198.6874 & -198.3752 & 0.3122 & -198.3535 & 0.3339 & -198.6835 & 0.0039 & -198.6641 & 0.0233 \\ 
		& 1.35 & -198.7375 & -198.4445 & 0.2930 & -198.4006 & 0.3369 & -198.7346 & 0.0029 & -198.6902 & 0.0473 \\ 
		& 1.50 & -198.7497 & -198.4696 & 0.2801 & -198.3982 & 0.3515 & -198.7471 & 0.0026 & -198.6759 & 0.0738 \\ 
		& 1.65 & -198.7472 & -198.4784 & 0.2688 & -198.4650 & 0.2823 & -198.7467 & 0.0005 & -198.7370 & 0.0102 \\ 
		& 1.80 & -198.7403 & -198.4787 & 0.2616 & -198.4644 & 0.2759 & -198.7413 & 0.0010 & -198.7263 & 0.0141 \\ 
		& 1.95 & -198.7334 & -198.4758 & 0.2577 & -198.4601 & 0.2733 & -198.7349 & 0.0014 & -198.7167 & 0.0168 \\ 
		& 2.10 & -198.7282 & -198.4709 & 0.2573 & -198.4542 & 0.2740 & -198.7275 & 0.0007 & -198.7058 & 0.0224 \\ 
		& 2.25 & -198.7249 & -198.4684 & 0.2564 & -198.4544 & 0.2705 & -198.7247 & 0.0002 & -198.7018 & 0.0231 \\ 
		& 2.40 & -198.7232 & -198.4707 & 0.2525 & -198.4536 & 0.2696 & -198.7241 & 0.0009 & -198.6992 & 0.0240 \\ 
		& 2.55 & -198.7225 & -198.4678 & 0.2548 & -198.4495 & 0.2731 & -198.7218 & 0.0008 & -198.6908 & 0.0318 \\ 
		& 2.70 & -198.7225 & -198.4690 & 0.2535 & -198.4538 & 0.2686 & -198.7229 & 0.0004 & -198.6940 & 0.0285 \\ 
		
		\hline
		\multirow{12}{*}{N$_2$} 
		& 0.80 & -108.5079 & -107.3873 & 1.1206 & -107.8045 & 0.7034 & -108.5006 & 0.0073 & -108.5001 & 0.0078 \\ 
		& 0.95 & -108.9562 & -107.9841 & 0.9721 & -108.3324 & 0.6238 & -108.9345 & 0.0217 & -108.9367 & 0.0195 \\ 
		& 1.10 & -109.0415 & -108.0797 & 0.9619 & -108.4170 & 0.6245 & -109.0197 & 0.0218 & -108.9870 & 0.0545 \\ 
		& 1.25 & -108.9958 & -108.0769 & 0.9189 & -108.3955 & 0.6003 & -108.9397 & 0.0560 & -108.9155 & 0.0803 \\ 
		& 1.40 & -108.9205 & -108.0720 & 0.8485 & -108.3266 & 0.5939 & -108.8791 & 0.0414 & -108.7967 & 0.1238 \\ 
		& 1.55 & -108.8546 & -108.0669 & 0.7876 & -108.3015 & 0.5530 & -108.8119 & 0.0426 & -108.7245 & 0.1301 \\ 
		& 1.70 & -108.8017 & -108.0256 & 0.7761 & -108.2418 & 0.5599 & -108.7287 & 0.0729 & -108.6293 & 0.1724 \\ 
		& 1.85 & -108.7657 & -107.9901 & 0.7755 & -108.1791 & 0.5865 & -108.6584 & 0.1073 & -108.5380 & 0.2276 \\ 
		& 2.00 & -108.7468 & -107.9817 & 0.7651 & -108.1432 & 0.6036 & -108.6170 & 0.1298 & -108.4788 & 0.2680 \\ 
		& 2.15 & -108.7398 & -107.9877 & 0.7521 & -108.1047 & 0.6351 & -108.5871 & 0.1528 & -108.4215 & 0.3184 \\ 
		& 2.30 & -108.7385 & -108.0094 & 0.7291 & -108.1073 & 0.6312 & -108.5975 & 0.1410 & -108.4103 & 0.3282 \\ 
		& 2.45 & -108.7391 & -108.0175 & 0.7216 & -108.1244 & 0.6147 & -108.5954 & 0.1437 & -108.4159 & 0.3232 \\ 
		\hline\hline
	\end{tabular}
\end{table*}

\section{Convergence evaluation \label{appendix3}}
\begin{figure*}
	\centering
	\includegraphics[width=0.9\textwidth]{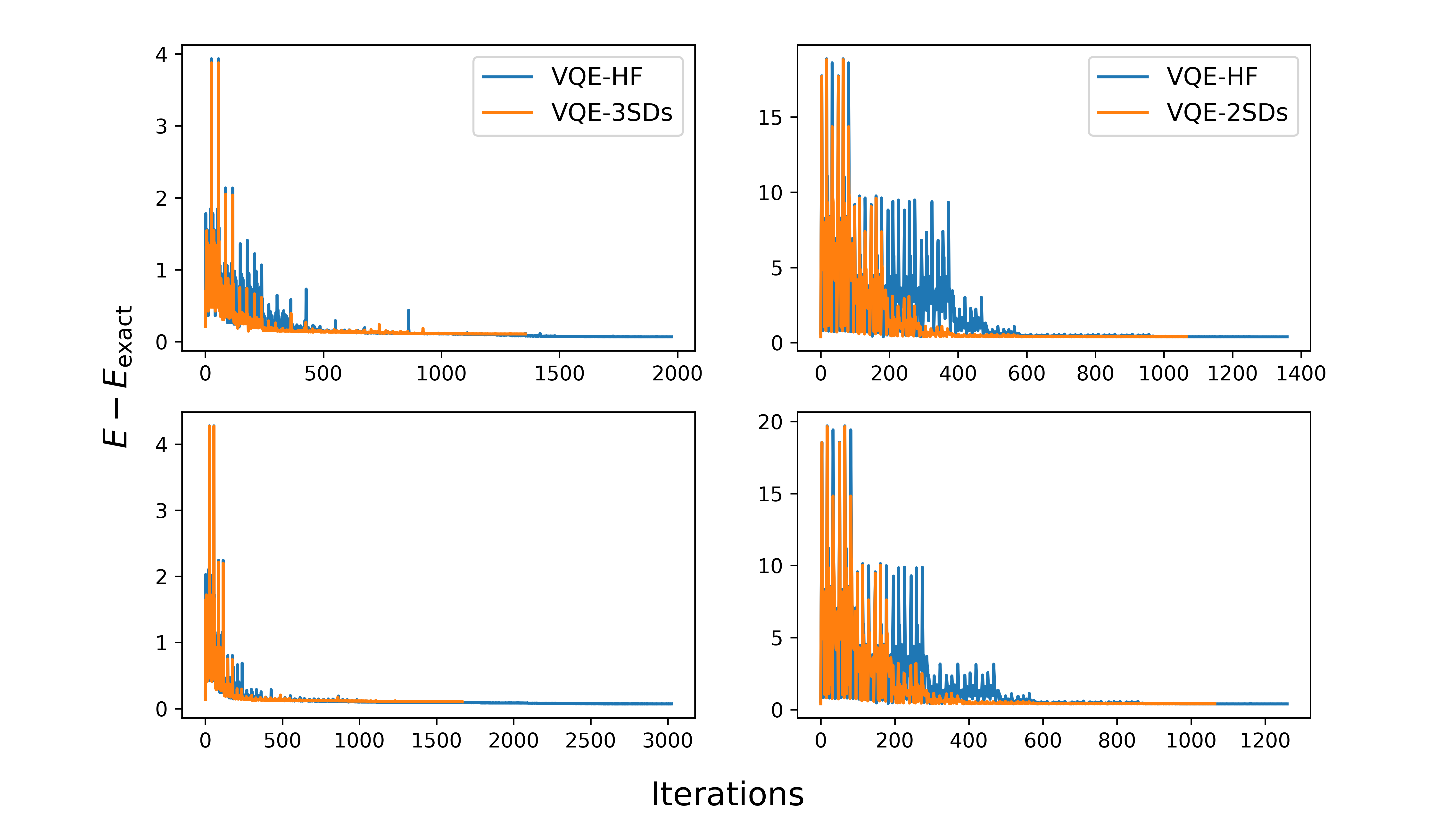}
	\caption{Convergence iterations of the VQE algorithm under the ImFil optimizer for four example systems. Left: $\mathrm{H_2O~(4e, 4o)}$ at $\mathrm{R=2.45 ~\AA}$ (top) and  $\mathrm{R=1.85 ~\AA}$ (bottom), and Right: $\mathrm{F_2~(10e, 6o)}$ at $\mathrm{R=2.7 ~\AA}$ (top) and $\mathrm{R=2.1 ~\AA}$ (bottom).}
	\label{converge}
\end{figure*}

MREM utilizes improved MR initial points, resulting in fewer iterations required for convergence in the VQE algorithm, as illustrated in several examples in Fig.~\ref{converge}.

\end{document}